\documentclass[twocolumn,prl]{revtex4}
\usepackage{graphicx,epsfig,amssymb}
\usepackage{amsmath}
\usepackage{graphicx}
\usepackage{amsfonts}
\usepackage{amssymb}

\begin{document}


\maketitle

{\bf Feldbacher, Held, and Assaad Reply:} 
Recently, we introduced a
projective quantum Monte Carlo (PQMC) algorithm  for simulating the
Anderson impurity model (AIM) \cite{Feldbacher}.
The  preceding critique \cite{Katsnelson} based on the
 orthogonality catastrophe (OC) \cite{Anderson} is invalid:
(i) There is no OC
in \cite{Feldbacher}, and
 it is generally not
 ``unpractical'' to avoid it.
 (ii) The OC does not affect our
results.

Concerning  (i): The OC theorem  \cite{Anderson} states: If a Hamiltonian $H_T$ is perturbed by a
local disorder potential and/or interaction $H'$ to a Hamiltonian $H=H_T+H'$, the overlap of the ground states of $H$ and $H_T$ is
\begin{equation}
           | \langle \psi_G | \psi_T \rangle| \sim N^{-\alpha}, N \mbox{being the numer of bath sites}.
\label{Eq:OC} 
\end{equation}
What
Katsnelson \cite{Katsnelson} overlooked is that $\alpha=0$ in
 \cite{Feldbacher}. The  $\alpha$ of Eq.\ (\ref{Eq:OC})
is given by the difference between the 
scattering phase for
$H_T$ and $H$ ($\varphi_T$ and $\varphi_G$, respectively), i.e.,
$\alpha \sim (\varphi_T-\varphi_G)^2$\cite{Yamada}.
According to the Friedel sum rule, the scattering phase $\varphi$ is
related to the average number of electrons on the impurity site:  $\varphi = {\pi}/{2} \;  n_d$.
Since  $n_d$  is the same for $H$ and $H'$, 
we have $\varphi_T = \varphi_G$,
and hence $\alpha = 0$. There is no OC.

Note that our calculations in \cite{Feldbacher,Arita1} are 
for half-filled bands, so that we
automatically have $\alpha=0$. But also off half-filling one
is free to choose a $H_T$ with the 
same $n_d$. 

This proof based on Friedel's sum
rule has been incorporated in 
 \cite{Katsnelson}, where it is now argued
 that using a  $H_T$ without OC
is ``unpractical''  \cite{Katsnelson}
because one would need an (analytically) ``exact 
answer for $n_d$'' \cite{Katsnelson}.
We reject this critique.
For a {\em numerical} algorithm, it is
perfectly legitimate to calculate  $n_d$ {\em numerically}.
We have done so in practice  \cite{Arita}.

Concerning (ii):
The OC is irrelevant for our calculations since 
for long enough projection time $\theta$ we obtain
the same Green function $G(\tau)$ with and without OC, see Fig.\ 1. 
\begin{figure}[t]
\epsfxsize=7.2cm \epsfbox{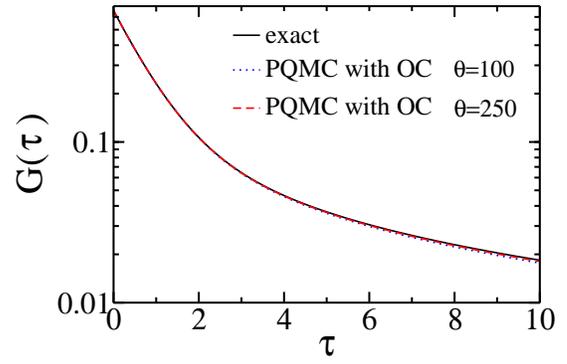}
\caption{(Color online) PQMC Green function with OC  and without OC (exact).
For long enough projection $\theta$,
the Green function with OC converges to
the exact result.
Note  the logarithmic scale of the $y$ axis.}
\end{figure}
The calculations of  Fig.\ 1 have been done for an AIM with 
on-site hybridization $V\!=\!1$, a constant bath density of states
ranging from -1 to 1 (with $N\rightarrow\infty$ bath sites), imaginary time discretization  $\Delta\tau\!=\!0.1$, impurity level $\epsilon_d\!=\!0.5$,
and Coulomb interaction $U=0$ (for maximizing precision).  The trial AIM  with OC was 
 half-filled, i.e., $\epsilon_d^{\rm trial}\!=\!0$, $n_d^{\rm trial}\!=\!1$
instead of  $\epsilon_d\!=\!0.5$, $n_d\!\sim\!0.7$. 
For long enough projection $\theta$,
the Green function with OC converges to
the exact result.  At $\theta\!=\!100$ (250), the average difference to the exact $G(\tau)$ is already as small as $7 \cdot 10^{-4}$ ($3 \cdot 10^{-4}$). Also,  $n_d=2[1 - G(\tau = 0)]$ is obtained
extremely precisely even with OC.

Technically, one can understand the irrelevance of the OC
by 
considering that we have integrated out the $N$ bath sites
of Eq.\ (\ref{Eq:OC}) \cite{HirschFye,Feldbacher}.
The algorithm is exactly the same with OC ($N\rightarrow\infty$) and without OC (finite $N$).
For large enough but finite $N$, also the  non-interacting input Green function 
can be the same as for $N\rightarrow \infty$, within any given accuracy. Hence, the results are the
same with and without OC.

Physically, one can understand the irrelevance of the OC
 by considering that the
OC theorem \cite{Anderson} holds in the
metallic, (quasi-)particle regime and only for $N\rightarrow \infty$.
Then, however, the ground state and
 excited states 
with a finite number of particle-hole excitations,
have the same physical properties, since
the finite number of particle-hole
excitations becomes irrelevant for $N\rightarrow\infty$.
Ground state and such low-lying excited states 
 yield the 
same equilibrium (ground state) Green function \cite{GS}. This explains, why we
obtain the correct results, e.g., in Fig.\ 1, even with OC.
The OC is irrelevant for our calculations. Katsnelson's 
objection \cite {Katsnelson} is not valid.

\noindent
M.\ Feldbacher$^1$, K.\ Held$^1$, and F.\ F.\ Assaad$^2$\\
{\small $^1$Max-Planck-Institut f\"ur Festk\"orperforschung, Heisenbergstra{\ss}e 1, D-70569 Stuttgart, Germany}\\
{\small  $^2$ Institut f\"ur theoretische Physik und Astrophysik, Universit\"at W\"urzburg, Am Hubland, D-97074 W\"urzburg, Germany}

{\small \noindent
Received\\
DOI:\\
PACS numbers:}


\begin{thebibliography}{99}

\bibitem{Feldbacher} M.\ Feldbacher, K.\ Held, and F.\ F.\ Assaad, Phys.\ Rev.\ Lett.\ {\bf 93}, 136405 (2004).

\bibitem{Katsnelson} M.\ I.\ Katsnelson, preceding comment.


\bibitem{Anderson} P.\ W.\ Anderson, Phys.\ Rev.\ Lett.\ {\bf 18}, 1049 (1967);
Phys.\ Rev.\ {\bf 164}, 352 (1967).

\bibitem{Yamada}
See, e.g.,
K.\ Yamada and K.\ Yosida, Progr. Theor. Phys. {\bf 60}, 353 (1978).







\bibitem{Arita1} R.\ Arita and K.\ Held, Phys. Rev. B {\bf 72}, 201102 (2005).


\bibitem{Arita} R.\ Arita and K.\ Held, Phys. Rev.  B   {\bf 73}, 064515 (2006).

\bibitem{HirschFye} J.\ E.\ Hirsch and R.\ M.\ Fye,
Phys.\ Rev.\ Lett.\ \textbf{56}, 2521 (1986).


\bibitem{GS} 
Also note 
O.\ Gunnarsson and K.\ Sch\"onhammer, Phys.\ Rev.\ B {\bf 26}, 2765 (1982),
where---for  the original OC model
of non-interacting electrons \cite{Anderson}---it was proved that
  the ground state and
the lowest excited states (even with infinitely many
particle-hole excitations,  less then $\sim N^{1/2}$)
yield the same local
expectation values for equal-time one-particle
operators in the limit $N\rightarrow\infty$.




\end{thebibliography}
\end{document}